\documentclass{article}

\usepackage[dvips]{graphicx}  

\begin{document}

\title{\LARGE \bf Dynamical  Fractal  3-Space and the Generalised Schr\"{o}dinger  Equation: Equivalence
Principle and  Vorticity Effects }  
\author{Reginald T. Cahill\\
{\it \small School of Chemistry, Physics and Earth Sciences}\\
 {\it \small Flinders University} \\ 
{\it \small GPO Box 2100, Adelaide 5001, Australia} \\
{\it \small(Reg.Cahill@flinders.edu.au)}}

\date{}
\maketitle

\begin{center}

{\small\parbox{11cm}{%
 The new dynamical `quantum foam' theory of 3-space is described at the classical 
level by a velocity field. This has been repeatedly  detected and for which the dynamical
equations are now established. These equations predict 3-space `gravitational wave' effects, and these have been
observed, and the 1991 DeWitte data is analysed to reveal the fractal structure of these
`gravitational waves'. This velocity field describes the differential motion of 3-space, and the various
equations of physics must be generalised to incorporate this 3-space dynamics.  Here a new  generalised
Schr\"{o}dinger  equation is given and analysed. It is shown that from this equation the equivalence principle
may be derived as a quantum effect, and that as well this generalised Schr\"{o}dinger equation determines the
effects of vorticity of the 3-space flow, or `frame-dragging', on matter, and which is being studied by the
Gravity Probe B (GP-B) satellite gyroscope experiment. 
\rule[0pt]{0pt}{0pt}}}\bigskip

\end{center}

\section{Introduction\label{section:introduction}}

Extensive experimental evidence  \cite{Book,AMGE,MM} has shown that a complex dynamical 3-space
underlies reality.  The evidence involves the repeated detection of the  motion of the earth
relative to that 3-space using Michelson interferometers  operating in gas mode \cite{MM},
particularly the experiment by Miller  \cite{Miller} in 1925/26 at Mt.Wilson, and the coaxial cable  RF travel
time measurements by Torr and Kolen in Utah in 1981, and the DeWitte experiment in 1991 in Brussels
\cite{MM}.  All such 7 experiments are consistent with respect to speed and direction. It has
been shown that
  effects caused by motion relative to this 3-space can mimic the formalism of
spacetime, but that it is the 3-space that is `real', simply because it is directly
observable \cite{Book}. 

The 3-space is in differential motion, that is one part has a velocity relative
to  other parts, and so involves a velocity field ${\bf v}({\bf r},t)$ description. To be
specific this velocity field must be described relative to a frame of observers, but the
formalism is such that the dynamical equations for this velocity field must transform
covariantly under a change of observer. As shown herein the experimental data from the DeWitte experiment
shows that  ${\bf v}({\bf r},t)$ has  a fractal structure. This arises because, in the absence of
matter, the dynamical equations for  ${\bf v}({\bf r},t)$ have no scale. This implies that the
differential motion of 3-space manifests at all scales.  This fractal differential motion of
3-space is missing from all the fundamental equations of physics, and so these equations
require a generalisation.  Here we report on the necessary generalisation of the
Schr\"{o}dinger equation, and which results in some remarkable results: (i) the equivalence
principle emerges, as well as (ii) the effects of vorticity of  this velocity field. These two effects are thus
seen to be quantum-theoretic effects, i.e. consequences of the wave nature of matter. The equivalence principle,
 as originally formulated by Galileo and then Newton, asserts that the gravitational acceleration of an object is
independent of its composition and speed.  However we shall see that via the vorticity effect, the velocity of
the object does affect the acceleration by causing rotations.

 It has been shown   \cite{Book,DMtrends} that the phenomenon of gravity is a consequence of the time-dependence
and inhomogeneities of  ${\bf v}({\bf r},t)$.  So the dynamical equations for  ${\bf v}({\bf
r},t)$ give rise to a new theory of gravity, when combined with the generalised Schr\"{o}dinger equation, and
the generalised Maxwell and Dirac equations. The equations for  ${\bf v}({\bf
r},t)$ involve the Newtonian gravitational constant $G$ and a dimensionless constant that determines the strength
of a new spatial self-interaction effect, which is missing from both Newtonian Gravity and General Relativity.
Experimental data has revealed
\cite{Book, DMtrends}  the remarkable discovery that this constant is the fine structure constant $\alpha
\approx 1/137$.  This dynamics then explains numerous gravitational anomalies, such as the bore hole $g$
anomaly, the so-called `dark matter' anomaly in the rotation speeds of spiral galaxies, and that the effective
mass of the necessary black holes at the centre of spherical matter systems, such as globular
clusters and spherical galaxies, is $\alpha/2$ times the total mass of these systems. This
prediction has been confirmed by astronomical  observations \cite{BH}.

The occurrence of $\alpha$ suggests that space is itself a quantum system undergoing on-going
classicalisation.  Just such a proposal has arisen in {\it Process Physics} \cite{Book} which is
an information-theoretic modelling of reality. There quantum space and matter arise in terms of
the Quantum Homotopic Field Theory (QHFT)  which, in turn, may be related to the standard model of
matter. In the QHFT space at this quantum level is best described as a `quantum foam'. So 
 we  interpret the observed fractal 3-space as a classical approximation
to this `quantum foam'. 

While here we investigate the properties of the generalised Schr\"{o}dinger
equation, analogous generalisations of the Maxwell  and Dirac equations, and in turn the
corresponding generalisations to the quantum field theories for such systems, may also be made. 
In the case of the Maxwell equations we obtain the light bending effects, including in particular gravitational
lensing, caused by the 3-space
 differential and time-dependent flow.

\section{The Physics of 3-Space\label{section:space}}

Because of the dominance of the spacetime ontology, which has been the
foundation of physics over the last century, the existence of a 3-space as an observable
phenomenon has been overlooked, despite extensive experimental detection over that period, and
earlier. This spacetime ontology is  distinct from the role of spacetime as a mathematical
formalism implicitly incorporating some real dynamical effects, though this distinction is rarely made. 
Consequently the existence of 3-space has been denied, and so there has never been a dynamical theory
for 3-space.  In recent years this situation has dramatically changed. We briefly
summarise the key aspects to the dynamics of 3-space. 

Relative to some observer 3-space is described by a velocity field  ${\bf v}({\bf r},t)$.
It is important to note that the coordinate ${\bf r}$  is not itself 3-space, rather it
is merely a label for an element of 3-space that has  velocity ${\bf v}$, relative to some
observer. This will become more evident when we consider the necessary generalisation of the
Schr\"{o}dinger equation. Also it is important to appreciate that this `moving' 3-space is not
itself  embedded in a `space'; the 3-space is all there is, although as noted above its deeper
structure is that of a `quantum foam'.  

In the case of zero vorticity $\nabla\times{\bf v}={\bf 0}$ the 3-space dynamics is given by, in
the non-relativistic limit,
\begin{equation}
\nabla.\left(\frac{\partial {\bf v} }{\partial t}+({\bf v}.{\bf \nabla}){\bf v}\right)
+\frac{\alpha}{8}\left((tr D)^2 - tr(D^2)\right)=
-4\pi G\rho,
\label{eqn:E1}\end{equation}
where $\rho$ is the matter density, and where 
\begin{equation} D_{ij}=\frac{1}{2}\left(\frac{\partial v_i}{\partial x_j}+
\frac{\partial v_j}{\partial x_i}\right).
\label{eqn:E2}\end{equation}
The acceleration of an element of space is given by the Euler form
\begin{eqnarray}
{\bf g}({\bf r},t)&\equiv&\lim_{\Delta t \rightarrow 0}\frac{{\bf v}({\bf r}+{\bf v}({\bf r},t)\Delta t,t+\Delta
t)-{\bf v}({\bf r},t)}{\Delta t} \nonumber \\
&=&\frac{\partial {\bf v}}{\partial t}+({\bf v}.\nabla ){\bf v}
\label{eqn:E3}\end{eqnarray} 
These forms are mandated by Galilean covariance under change of observer\footnote{However this does
not exclude  so-called relativistic effects, such as the length contraction of moving rods or the time
dilations of moving clocks.}.   This  non-relativistic modelling of the dynamics for the
velocity field gives a direct account of the various phenomena noted above. A generalisation to include
vorticity and  relativistic effects of the motion of matter through this 3-space is given in \cite{Book}.
From (\ref{eqn:E1}) and (\ref{eqn:E2}) we obtain that
\begin{equation}
\nabla.{\bf g}=-4\pi G\rho-4\pi G \rho_{DM},
\label{eqn:E4}\end{equation}
where 
\begin{equation}
\rho_{DM}({\bf r})=\frac{\alpha}{32\pi G}( (tr D)^2-tr(D^2)).  
\label{eqn:E5}\end{equation}
In this form we see that if $\alpha\rightarrow 0$, then the acceleration of the 3-space elements is given
by Newton's Law of Gravitation, in differential form. But for a non-zero $\alpha$ we see that the
3-space acceleration has an additional effect, the $\rho_{DM}$ term, which is an effective `matter density' that
mimics the new self-interaction dynamics.  This has been shown to be the origin of the so-called `dark
matter' effect in spiral galaxies.  It is important to note that (\ref{eqn:E4})  does not determine ${\bf
g}$ directly; rather the velocity dynamics in (\ref{eqn:E1}) must be solved, and then with ${\bf g}$
subsequently determined from (\ref{eqn:E3}).  Eqn.(\ref{eqn:E4}) merely indicates that the resultant
non-Newtonian aspects to ${\bf g}$  could be mistaken as being the result of a new form of matter, whose density
is given by $\rho_{DM}$. Of course the saga of `dark matter' shows that this actually happened, and that there
has been a misguided and fruitless search for such `matter'.  

The numerous experimental confirmations of (\ref{eqn:E1}) imply that Newtonian gravity is not universal
at all. Rather a key aspect to gravity was missed by Newton because it so happens that the
3-space self-interaction  dynamics does not necessarily explicitly manifest outside of spherical matter systems,
such as the sun. To see this it is only necessary to see that the velocity field 
\begin{equation}
{\bf v}({\bf r})=-\sqrt{\frac{2GM'}{r}}\hat{\bf r},
\label{eqn:E6}\end{equation} 
is a solution to (\ref{eqn:E1}) external to a spherical mass $M$, where $M'=(1+\frac{\alpha}{2})M+..$.
Then (\ref{eqn:E6}) gives, using (\ref{eqn:E3}), the resultant external `inverse square law' acceleration
\begin{equation}
{\bf g}({\bf r})=-\frac{GM'}{r^2}\hat{\bf r}.
\label{eqn:E7}\end{equation} 
Hence in this special case the 3-space dynamics predicts an inverse square law form for ${\bf g}$, as
confirmed in the non-relativistic regime by Kepler's laws for planetary motion, with only a modified value for
the effective mass $M'$. So for this reason we see how easy  it was for Newton to have overlooked a velocity
formalism for gravity, and so missed the self-interaction dynamics in (\ref{eqn:E1}).  Inside a spherical matter
system Newtonian gravity and the new gravity theory differ, and it was this difference that explained the bore
hole $g$ anomaly data
\cite{DMtrends}, namely that $g$ does not decrease down a bore hole as rapidly as Newtonian gravity predicts. It
was this anomaly that lead to the discovery that $\alpha$ was in fact the fine structure constant, up to
experimental errors.
As well the 3-space dynamics in (\ref{eqn:E1}) has `gravitational wave'  solutions \cite{QFGGW}. Then there
are regions where the velocity differs slightly from the enveloping region.  In the absence of matter these
waves will be in general fractal because there is no dimensioned constant, and so no natural scale.  These waves
were seen by Miller, Torr and Kolen, and by DeWitte \cite{Book,QFGGW} as shown in Fig.\ref{fig:fractal}.

However an assumption made in previous analyses was that the acceleration of the 3-space itself, in
(\ref{eqn:E3}), was also the acceleration of matter located in that 3-space.  The key result herein is to
derive this result by using the generalised Schr\"{o}dinger equation. In doing so we discover the additional
effect that vorticity in the velocity field causes quantum states to be rotated, as discussed in
Sect.\ref{section:GPB}.

\section{Newtonian Gravity and the  Schr\"{o}dinger Equation\label{section:newtonian}}

Let us consider what might be regarded as the conventional `Newtonian' approach to including gravity in the
Schr\"{o}dinger equation \cite{Schrod}.  There gravity is described by the Newtonian potential energy field
$\Phi({\bf r},t)$, such that ${\bf g}=-\nabla \Phi$, and we have for a `free-falling' quantum system, with mass
$m$,  
\begin{equation}
i\hbar\frac{\partial \psi({\bf r},t)}{\partial t}=-\frac{\hbar^2}{2m}\nabla^2\psi({\bf r},t)+
m\Phi({\bf r},t)\psi({\bf r},t)\equiv H(t)\Psi, 
\label{eqn:equiv1}\end{equation}
where the hamiltonian is in general now time dependent, because the masses producing the gravitational acceleration
may be moving. Then  the classical-limit trajectory  is obtained via the usual Ehrenfest method \cite{Ehrenfest}:
we first compute the time rate of change of the  so-called position `expectation value'
\begin{eqnarray}
 \frac{d\!\!<\!\!{\bf r}\!\!> }{dt} &\equiv& \frac{d }{dt}(\psi,{\bf
r}\psi)=\frac{i}{\hbar}(H\psi,{\bf r}\psi)-\frac{i}{\hbar}(\psi,{\bf r}H\psi)\nonumber
\\&=&\frac{i}{\hbar}(\psi,[H,{\bf r}]\psi),
\label{eqn:equiv2}\end{eqnarray}
which is valid for a normalised state $\psi$.  The norm is time invariant when $H$ is hermitian
($H^\dagger=H$) even if $H$  itself is time dependent,
\begin{eqnarray}
\frac{d}{dt}(\psi,\psi)&=&\frac{i}{\hbar}(H\psi,\psi)-\frac{i}{\hbar}(\psi,H\psi)\nonumber \\&=&
\frac{i}{\hbar}(\psi,H^\dagger\psi)-\frac{i}{\hbar}(\psi,H\psi)=0.
\label{eqn:equiv3}\end{eqnarray}
Next we compute the matter `acceleration' from (\ref{eqn:equiv2}).
\begin{eqnarray}
\frac{d^2\!\!<\!\!{\bf r}\!\!> }{dt^2} &=& \frac{i}{\hbar}\frac{d }{dt}(\psi,[H,{\bf r}]\psi),\nonumber\\ 
&=&\left(\frac{i}{\hbar}\right)^2(\psi,[H,[H,{\bf r}]]\psi)+\frac{i}{\hbar}(\psi,[\frac{\partial H(t)
}{\partial t},{\bf r}]\psi),\nonumber\\
&=&-(\psi,\nabla \Phi\psi)=(\psi,{\bf g}({\bf r},t)\psi)=<\!{\bf g}({\bf r},t)\!>.
\label{eqn:equiv4}\end{eqnarray}
where  for the commutator
\begin{equation}\left[\frac{\partial H(t)}{\partial t},{\bf r}\right]=\left[m\frac{\partial \Phi({\bf r}, t)}{\partial
t},{\bf r}\right]=0.\end{equation}
 In the classical limit
$\psi$ has the form of a wavepacket where the spatial extent of $\psi$ is much smaller than the spatial region over
which  ${\bf g}({\bf r},t)$ varies appreciably.  Then we have the approximation $<\!{\bf g}({\bf r},t)\!>\approx {\bf
g}(<\!{\bf r}\!>,t)$, and finally we arrive at the Newtonian 2nd-law equation of motion for the wavepacket,
 \begin{equation}
\frac{d^2\!\!<\!\!{\bf r}\!\!> }{dt^2}\approx {\bf g}(<\!{\bf r}\!>, t).
\label{eqn:equiv5}\end{equation}
 In this classical
limit we obtain the equivalence principle, namely that the acceleration is independent of the mass $m$ and of
the velocity of that mass. But of course that followed by construction, as the equivalence principle is built
into (\ref{eqn:equiv1}) by having
$m$ as the coefficient of $\Phi$. In Newtonian gravity there is no explanation for the origin of $\Phi$ or
${\bf g}$. In the new theory gravity is explained in terms of a velocity field, which in turn has a deeper
explanation within {\it Process Physics}.

\section{Dynamical 3-Space and  the Generalised   Schr\"{o}dinger
Equation\label{section:schrodinger}}

The key insight is that conventional physics has neglected the interaction of various systems with the dynamical
3-space.  Here we  generalise the Schr\"{o}dinger equation to take account of this new physics. Now gravity is a
dynamical effect arising from the time-dependence and spatial inhomogeneities of the 3-space  velocity field ${\bf
v}({\bf r},t)$, and  for a `free-falling' quantum system with mass
$m$ the Schr\"{o}dinger equation now has the generalised form  
\begin{equation}
i\hbar\left(\frac{\partial}{\partial t} +{\bf
v}.\nabla+\frac{1}{2}\nabla.{\bf v}\right) \psi({\bf r},t)=-\frac{\hbar^2}{2m}\nabla^2\psi({\bf r},t), 
\label{eqn:equiv6}\end{equation}
which we write as 
\begin{equation}
i\hbar\frac{\partial  \psi({\bf r},t)}{\partial t}=H(t)\psi({\bf r},t),
\label{eqn:equiv7}\end{equation}
where now
\begin{equation}
H(t)=-i\hbar\left({\bf
v}.\nabla+\frac{1}{2}\nabla.{\bf v}\right)-\frac{\hbar^2}{2m}\nabla^2
\label{eqn:equiv8}\end{equation}
This form for $H$   specifies how the quantum system must couple to the velocity field, and it  uniquely 
follows from two considerations: (i) the generalised Schr\"{o}dinger equation must remain form invariant under a
change of observer, i.e. with $t
\rightarrow t$, and ${\bf r}
\rightarrow {\bf r}+{\bf V}t$, where ${\bf V}$ is the relative velocity of the two observers. Then we compute
that
$\displaystyle{\frac{\partial}{\partial t} +{\bf v}.\nabla +\frac{1}{2}\nabla.{\bf v} \rightarrow} $ $
\displaystyle{\frac{\partial}{\partial t} +{\bf v}.\nabla}+\frac{1}{2}\nabla.{\bf v}$, i.e. that it is an
invariant operator, and   (ii) requiring that
$H(t)$ be hermitian, so that the wavefunction norm is an invariant of the time evolution. This implies that the
$\frac{1}{2}\nabla.{\bf v}$ term must be included, as ${\bf v}.\nabla$ by itself is not hermitian for an
inhomogeneous ${\bf v}({\bf r},t)$. Then the consequences for the motion of wavepackets are uniquely determined;
they are fixed by these two quantum-theoretic requirements\footnote{For two or more `particles' we have by the
same arguments $H(t)=\sum_j-i\hbar\left({\bf
v}.\nabla_j+\frac{1}{2}\nabla_j.{\bf v}\right)-\frac{\hbar^2}{2m_j}\nabla^2_j$}.

Then  again the classical-limit trajectory  is obtained via the position `expectation value', first with
\begin{eqnarray}
{\bf v}_O\equiv\frac{d\!\!<\!\!{\bf r}\!\!> }{dt} &=& \frac{d }{dt}(\psi,{\bf
r}\psi)=\frac{i}{\hbar}(\psi,[H,{\bf r}]\psi),\nonumber \\ &=&(\psi,({\bf v}({\bf r},
t)-\frac{i\hbar}{m}\nabla)\psi)\nonumber \\ &=&<\!\!{\bf v}({\bf r}, t)\!\!>-\frac{i\hbar}{m}<\!\!\nabla\!\!>,
\label{eqn:equiv9}\end{eqnarray}
on evaluating the commutator using  $H(t)$ in (\ref{eqn:equiv8}), and which is again valid for a normalised state
$\psi$. 

Then for the `acceleration' we obtain from (\ref{eqn:equiv9})  that\footnote{Care is needed to indicate the
range of the various $\nabla$'s. Extra parentheses $($ ... $)$ are used to limit the range when required.}
\begin{eqnarray}
\lefteqn{\frac{d^2\!\!<\!\!{\bf r}\!\!> }{dt^2} = \frac{d }{dt}(\psi,({\bf v}
-\frac{i\hbar}{m}\nabla)\psi)}\nonumber\\ 
& & =(\psi,\left(\frac{\partial {\bf v}({\bf r},t) }{\partial
t}+\frac{i}{\hbar}[H,({\bf v} -\frac{i\hbar}{m}\nabla)]\right)\psi),\nonumber\\
& &=(\psi,\frac{\partial {\bf v}({\bf r},t) }{\partial t}\psi)+
(\psi,
\left({\bf
v}.\nabla+\frac{1}{2}\nabla.{\bf v}-\frac{i\hbar}{2m}\nabla^2\right)\left({\bf v}
-\frac{i\hbar}{m}\nabla\right)\psi)-\nonumber\\
& &\mbox{\ \ \ }(\psi,\left.\left({\bf v}
-\frac{i\hbar}{m}\nabla\right)\left({\bf v}.\nabla+\frac{1}{2}\nabla.{\bf
v}-\frac{i\hbar}{2m}\nabla^2\right)\right)\psi), \nonumber\\ 
& & =(\psi,\left(\frac{\partial {\bf
v}({\bf r},t) }{\partial t}+(({\bf v}.\nabla){\bf v}) -\frac{i\hbar}{m}(\nabla\times{\bf
v})\times {\bf \nabla}\right)\psi)+(\psi,\frac{i\hbar}{2m}(\nabla\times(\nabla\times {\bf v}))\psi),\nonumber \\
& &\approx\frac{\partial{\bf v}}{\partial t}+({\bf v}.\nabla){\bf v}+(\nabla\times{\bf
v})\times\left(\frac{d\!\!<\!\!{\bf r}\!\!> }{dt}-{\bf v}\right)+\frac{i\hbar}{2m}(\nabla\times(\nabla\times{\bf
v})),\nonumber \\ & &=\frac{\partial{\bf v}}{\partial t}+({\bf v}.\nabla){\bf v}+(\nabla\times{\bf
v})\times\left(\frac{d\!\!<\!\!{\bf r}\!\!> }{dt}-{\bf v}\right)\nonumber \\
& &=\frac{\partial{\bf v}}{\partial t}+({\bf v}.\nabla){\bf v}+
(\nabla\times{\bf v})\times{\bf v}_R
\label{eqn:equiv10}\end{eqnarray}
where in arriving at the 3rd last line we have invoked the small-wavepacket approximation, and 
also used (\ref{eqn:equiv9}) to identify 
\begin{equation} {\bf v}_R \equiv -\frac{i\hbar}{m}<\!\!\nabla\!\!>={\bf v}_O-{\bf v},
\label{eqn:equiv11}\end{equation}
where ${\bf v}_O$  is the velocity of the wavepacket or object `O' relative to the observer, so then ${\bf v}_R$
is the velocity of the wavepacket relative to the local 3-space. 
Then  all
velocity field terms are now evaluated at the location of the wavepacket. 
Note that the operator
\begin{equation}
-\frac{i\hbar}{m}(\nabla\times{\bf v})\times\nabla+\frac{i\hbar}{2m}(\nabla\times(\nabla\times{\bf v}))
\end{equation} is hermitian, but that separately neither of these two operators is hermitian. Then in general
the scalar product in  (\ref{eqn:equiv10}) is real. But then in arriving at the last line in (\ref{eqn:equiv10}) 
by means of the small-wavepacket approximation, we must then self-consistently use that 
$\nabla\times(\nabla\times{\bf v})={\bf 0}$, otherwise the acceleration acquires a spurious imaginary part.  
This is consistent with  (\ref{eqn:CG4b}) outside of any  matter which contributes to the generation of the
velocity field, for  there
$\rho=0$. These observations point to a deep connection between quantum theory and the velocity field dynamics,
as already  argued in \cite{Book}.

We see that the
test `particle' acquires the acceleration of the velocity field, as in (\ref{eqn:E3}), and as well  an
additional vorticity induced acceleration which is the analogue of the Helmholtz acceleration in fluid mechanics. 
Then $\vec{\omega}/2$ is the instantaneous angular velocity of the local 3-space, relative to a distant
observer. Hence we find that the equivalence principle arises from the unique generalised Schr\"{o}dinger
equation and with the additional vorticity effect.  This vorticity effect depends on the absolute velocity ${\bf
v}_R$ of the object relative to the local space, and so requires a change in the Galilean or Newtonian form of
the equivalence principle.

The vorticity acceleration effect is the origin of the
Lense-Thirring so-called `frame-dragging' \footnote{In the spacetime formalism it is mistakenly argued that it
is `spacetime' that is `dragged'.} effect
\cite{LT} discussed in Sect.\ref{section:GPB}. While the generation of the vorticity is a relativistic effect,
as in (\ref{eqn:CG4b}), the response of the  test particle to that vorticity is a non-relativistic effect, and
follows from the generalised Schr\"{o}dinger equation, and which is not present in the standard Schr\"{o}dinger
equation with coupling to the Newtonian gravitational potential, as in (\ref{eqn:equiv1}). Hence the generalised 
Schr\"{o}dinger equation with the new coupling to the velocity field is  more fundamental. The Helmholtz term in
(\ref{eqn:equiv10}) is being explored by the Gravity Probe B gyroscope precession experiment, however the
vorticity caused by the motion of the earth is extremely small, as discussed in Sect.\ref{section:GPB}. 

An important insight emerges from the form of (\ref{eqn:equiv7}) and (\ref{eqn:equiv8}): here the
generalised Schr\"{o}dinger equation involves two fields  ${\bf v}({\bf r},t)$ and  
$\psi({\bf r},t)$, where the coordinate ${\bf r}$ is merely a label to relate the two fields, and is
not itself the 3-space.  In particular while ${\bf r}$ may have the form of a Euclidean 3-geometry,
the space itself has time-dependence and inhomogeneities, and as well in the more general case will
exhibit vorticity $\omega=\nabla\times{\bf v}$.  Only in the unphysical case does the description
of the 3-space become identified with the coordinate system ${\bf r}$, and that is when the velocity
field  ${\bf v}({\bf r},t)$ becomes uniform and time independent. Then by a suitable choice of
observer we may put  ${\bf v}({\bf r},t)={\bf 0}$, and the generalised Schr\"{o}dinger equation 
reduces to the usual `free' Schr\"{o}dinger equation.    As we discuss
later the experimental evidence is that  ${\bf v}({\bf r},t)$ is fractal and so cannot be removed by
a change to a preferred observer.  Hence the generalised Schr\"{o}dinger equation in
(\ref{eqn:equiv7})-(\ref{eqn:equiv8}) is a major development for fundamental physics. Of course in general  other
non-3-space potential energy terms may be added to the RHS of (\ref{eqn:equiv8}). A prediction of this
new quantum theory, which also extends to a generalised Dirac equation, is that the fractal structure to space
implies that even at the scale of atoms etc there will be time-dependencies and inhomogeneities, and that these
will affect  transition rates  of quantum systems.  These effects are probably 
 those known as the Shnoll effects \cite{Shnoll1}.

\section{Free-Fall Minimum Proper-Time Trajectories\label{section:geodesic}}

The acceleration in (\ref{eqn:equiv10}) also arises from the following  argument, which is the
analogue of the Fermat least-time formalism.  Consider the elapsed time for a comoving clock
travelling with the test particle. Then taking account of the Lamour time-dilation effect that
time is given by 
\begin{equation}
\tau[{\bf r}_0]=\int dt \left(1-\frac{{\bf v}_R^2}{c^2}\right)^{1/2}
\label{eqn:G1}
\end{equation}  
with ${\bf v}_R$ given by (\ref{eqn:equiv11}) in terms of ${\bf v}_O$ and ${\bf v}$. Then this time effect
relates to the speed of the clock relative to the local 3-space, and that $c$ is the speed of light relative to
that local 3-space. We are using a relativistic treatment in (\ref{eqn:G1}) to demonstrate the generality of the
results\footnote{ A non-relativistic analysis may be alternatively pursued by first expanding (\ref{eqn:G1}) in
powers of $1/c^2$.}. Under a deformation of the trajectory 
\begin{equation}{\bf r}_0(t)
\rightarrow  {\bf r}_0(t) +\delta{\bf r}_0(t),
\mbox{\ \ }{\bf v}_0(t) \rightarrow  {\bf v}_0(t) +\displaystyle\frac{d\delta{\bf r}_0(t)}{dt},\end{equation}
  and then
\begin{equation}\label{eqn:G2}
{\bf v}({\bf r}_0(t)+\delta{\bf r}_0(t),t) ={\bf v}({\bf r}_0(t),t)+(\delta{\bf
r}_0(t).{\bf \nabla}) {\bf v}({\bf r}_0(t),t)+... 
\end{equation}
Evaluating the change in proper travel time to lowest order
\begin{eqnarray*}\label{eqn:G3}
\delta\tau&=&\tau[{\bf r}_0+\delta{\bf r}_0]-\tau[{\bf r}_0] +... \nonumber\\
&=&-\int dt \:\frac{1}{c^2}{\bf v}_R. \delta{\bf v}_R\left(1-\displaystyle{\frac{{\bf
v}_R^2}{c^2}}\right)^{-1/2}+...\nonumber\\
&=&\int dt\frac{1}{c^2}\displaystyle{\frac{{\bf
v}_R.(\delta{\bf r}_0.{\bf \nabla}){\bf v}-{\bf v}_R.\displaystyle{\frac{d(\delta{\bf
r}_0)}{dt}}}{\sqrt{1-\displaystyle{\frac{{\bf v}_R^2}{c^2}}}}}+...\nonumber\\ 
&=&\int dt \frac{1}{c^2}\left(\frac{{\bf v}_R.(\delta{\bf r}_0.{\bf \nabla}){\bf v}}{ 
\sqrt{1-\displaystyle{\frac{{\bf
v}_R^2}{c^2}}}}  +\delta{\bf r}_0.\frac{d}{dt} 
\frac{{\bf v}_R}{\sqrt{1-\displaystyle{\frac{{\bf
v}_R^2}{c^2}}}}\right)+...\nonumber\\
&=&\int dt\: \frac{1}{c^2}\delta{\bf r}_0\:.\left(\frac{({\bf v}_R.{\bf \nabla}){\bf v}+{\bf v}_R\times({\bf
\nabla}\times{\bf v})}{ 
\sqrt{1-\displaystyle{\frac{{\bf v}_R^2}{c^2}}}}+\frac{d}{dt} 
\frac{{\bf v}_R}{\sqrt{1-\displaystyle{\frac{{\bf
v}_R^2}{c^2}}}}\right)+...\nonumber \\
\end{eqnarray*}
  Hence a 
trajectory ${\bf r}_0(t)$ determined by $\delta \tau=0$ to $O(\delta{\bf r}_0(t)^2)$
satisfies 
\begin{equation}\label{eqn:G4}
\frac{d}{dt} 
\frac{{\bf v}_R}{\sqrt{1-\displaystyle{\frac{{\bf v}_R^2}{c^2}}}}=-\frac{({\bf
v}_R.{\bf \nabla}){\bf v}+{\bf v}_R\times({\bf
\nabla}\times{\bf v})}{ 
\sqrt{1-\displaystyle{\frac{{\bf v}_R^2}{c^2}}}}.
\label{eqn:vReqn}\end{equation}
   Substituting ${\bf
v}_R(t)={\bf v}_0(t)-{\bf v}({\bf r}_0(t),t)$ and using 
\begin{equation}\label{eqn:G5}
\frac{d{\bf v}({\bf r}_0(t),t)}{dt}=\frac{\partial {\bf v}}{\partial t}+({\bf v}_0.{\bf \nabla}){\bf
v},
\end{equation}
we obtain
\begin{equation}\label{eqn:CG6}
 \frac{d {\bf v}_0}{dt}=\displaystyle{\frac{\partial {\bf v}}{\partial t}}+({\bf v}.{\bf \nabla}){\bf
v}+({\bf \nabla}\times{\bf v})\times{\bf v}_R-\frac{{\bf
v}_R}{1-\displaystyle{\frac{{\bf v}_R^2}{c^2}}}
\frac{1}{2}\frac{d}{dt}\left(\frac{{\bf v}_R^2}{c^2}\right).
\end{equation}
Then in the low speed limit  $v_R \ll c $   we  may neglect the last term, and we obtain
(\ref{eqn:equiv10}). Hence we see a close relationship between the geodesic equation, known first from General
Relativity, and the 3-space generalisation of the Schr\"{o}dinger equation, at least in the non-relativistic limit.
So in the classical limit, i.e when the   wavepacket approximation is valid, the wavepacket trajectory
 is specified by the least propertime geodesic.    

The relativistic term in (\ref{eqn:CG6}) is responsible for the precession of
elliptical orbits and also for the event horizon effect. Hence the trajectory  in (\ref{eqn:equiv10})
is a non-relativistic minimum travel-time trajectory, which is Fermat's Principle.  The relativistic
term in (\ref{eqn:CG6}) will arise from a  generalised  Dirac equation which would then include the
dynamics of  3-space.

\begin{figure}[t]
\hspace{5mm}\includegraphics[scale=0.41]{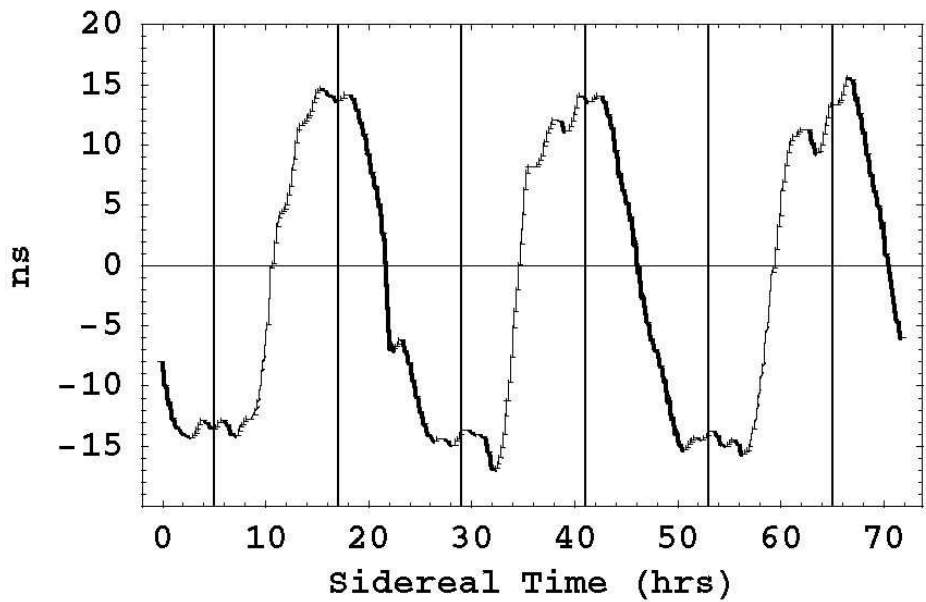}
\caption{\small{ Variations in twice the one-way travel time, in ns, for an RF signal to travel 1.5
km through a buried coaxial cable between  Rue du Marais and Rue de 
la Paille, Brussels. An offset  has been used  such that the average is zero.   The cable has a
North-South  orientation, and the data is $\pm$ difference of the travel times  for NS and SN
propagation.  The sidereal time for maximum  effect of $\sim\!\!5$hr (or   $\sim\!\!17$hr) (indicated
by vertical lines) agrees with the direction found by Miller \cite{Miller}. Plot shows
data over 3 sidereal days  and is plotted against sidereal time. The main effect is caused by the rotation of the
earth. The superimposed fluctuations are evidence of turbulence i.e gravitational
waves. Removing the earth induced rotation effect we obtain the first experimental data of the fractal
structure of space, and is shown in Fig.\ref{fig:fractal}. DeWitte performed this experiment over 178 days, and
demonstrated that the effect tracked sidereal time and not solar time \cite{Book}. }  
\label{fig:DeWittetimes}}\end{figure}

\begin{figure}[t]
\hspace{5mm}\includegraphics[scale=0.4]{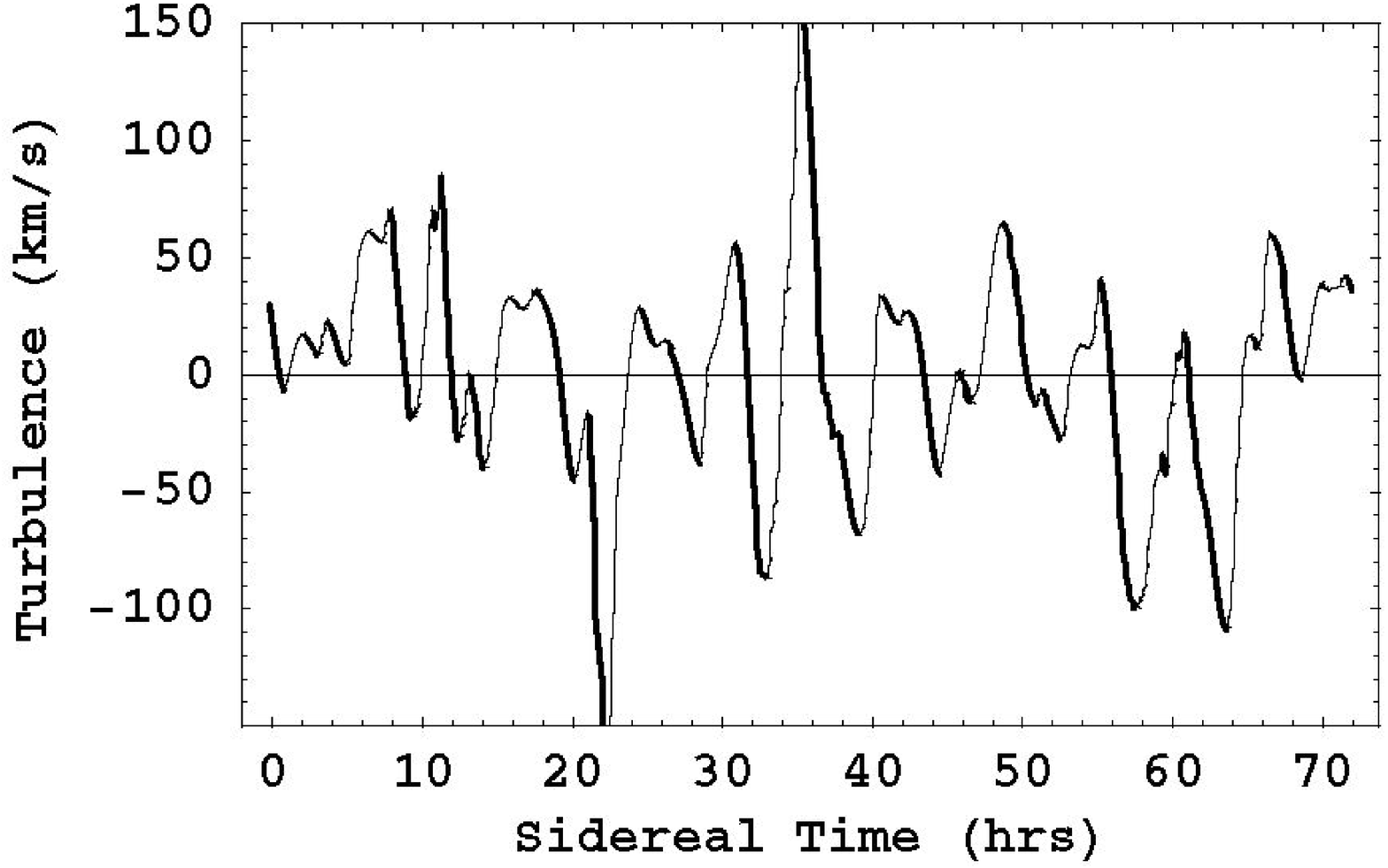}
\caption{\small{Shows the velocity fluctuations, essentially `gravitational waves' observed by 	DeWitte in 1991
from the measurement of variations in the RF coaxial-cable one-way travel times.  This data is obtained from that
in Fig.\ref{fig:DeWittetimes} after removal of the dominant effect caused by the rotation of the earth. 
Ideally the velocity fluctuations are three-dimensional, but the DeWitte experiment had only one 
arm. This plot is suggestive of a fractal structure to the velocity field. This is confirmed by the
power law analysis  shown in  Fig.\ref{fig:powerlaw}.}
\label{fig:fractal}}\end{figure}

\section{Fractal 3-Space and the DeWitte  Experimental Data\label{section:fractal}}

In 1991 Roland DeWitte working  within Belgacom,
the Belgium telecommunications company, accidently made yet another detection of absolute motion, and one
which was 1st-order in $v/c$.    5MHz radio frequency (RF) signals were sent  in both
directions   through two  buried  coaxial cables linking the two clusters of cesium atomic clocks.  

Changes in propagation times  were observed and eventually observations over  178 days were recorded. 
A sample of the  data, plotted against sidereal time for just  three days, is shown
in Fig.\ref{fig:DeWittetimes}. The  DeWitte  data was clear evidence of absolute motion with
the Right Ascension for minimum/maximum  propagation time agreeing almost exactly with Miller's direction
\footnote{This velocity arises after removing the effects of the earth's orbital  speed about the sun, 30km/s,
and the gravitational in-flow past the earth towards the sun, 42km/s, as in (\ref{eqn:E6}). } ($\alpha=5.2^{hr},
\delta=-67^0$)\footnote{The opposite direction is not easily excluded due to errors within the data, and so
should also be considered as possible.  A new experiment will be capable of more accurately determining the
speed and direction, as well as the fractal structure of 3-space. The author is constructung a more compact
version of the Torr-Kolen - DeWitte  coaxial cable RF travel-time experiment. New experimental techniques have
been developed to increase atomic-clock based timing accuracy and stability, so that shorter cables can be used,
which will permit 3-arm devices.}, and with speed
$420\pm 30$km/s. This local absolute motion is different from the CMB motion, in the direction
($\alpha=11.20^{hr},
\delta=-7.22^0$) with speed of
$369$km/s, for that would have given the data a totally different sidereal time signature, namely the times for
maximum/ minimum would have been shifted by 6hrs. The CMB velocity is motion relative to the distant early
universe, whereas the velocity measured in the DeWitte and related experiments is the velocity relative to the
local space. The declination of the velocity observed in this DeWitte experiment cannot be determined from the
data as only three days of data are available.   However assuming exactly the same declination as Miller  the
speed observed by DeWitte appears to be also  in excellent agreement with the Miller speed.   The dominant
effect in Fig.\ref{fig:DeWittetimes} is caused by the rotation of the earth, namely that the orientation of the
coaxial cable with respect to the direction of the flow past the earth changes as the earth rotates. This effect
may be approximately unfolded from the data, leaving  the gravitational waves shown in Fig.\ref{fig:fractal}. 
This is the first  evidence that the velocity field describing 3-space has a complex structure, and is indeed
fractal.

\begin{figure}[t]
\hspace{20mm}\includegraphics[scale=0.31]{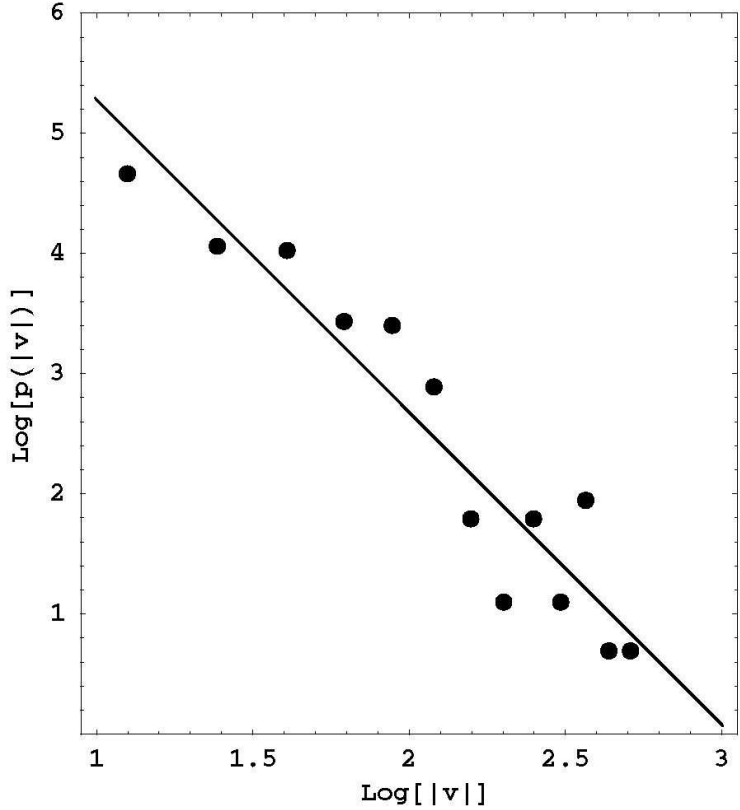}
\caption{\small{Shows that the velocity fluctuations in  Fig.\ref{fig:fractal} are scale free, as  the
probability distribution from binning the speeds has the form $p(v) \propto |v|^{-2.6}$. This plot shows
Log$[p(v)]$ vs Log$[|v|]$. This shows that
the velocity field has a fractal structure, and so requiring the generalisation of the Schr\"{o}dinger
equation, as discussed herein, and also the Maxwell and Dirac equations (to be discussed elsewhere).}
\label{fig:powerlaw}}\end{figure}

The fractal structure, i.e. that there is an
intrinsic lack of scale, to these speed fluctuations is demonstrated by binning the absolute speeds $|v|$
 and counting the number of speeds $p(|v|)$ within each bin. A least squares fit of the Log-Log plot to a
straightline was then made. Plotting  Log$[p(|v|)]$ vs
Log$[|v|]$, as shown  in Fig.\ref{fig:powerlaw}, we see that the fit gives
 $p(v) \propto |v|^{-2.6}$.   With the new experiment considerably more data will become available.

\begin{figure}[t]
\hspace{7mm}\includegraphics[scale=1.6]{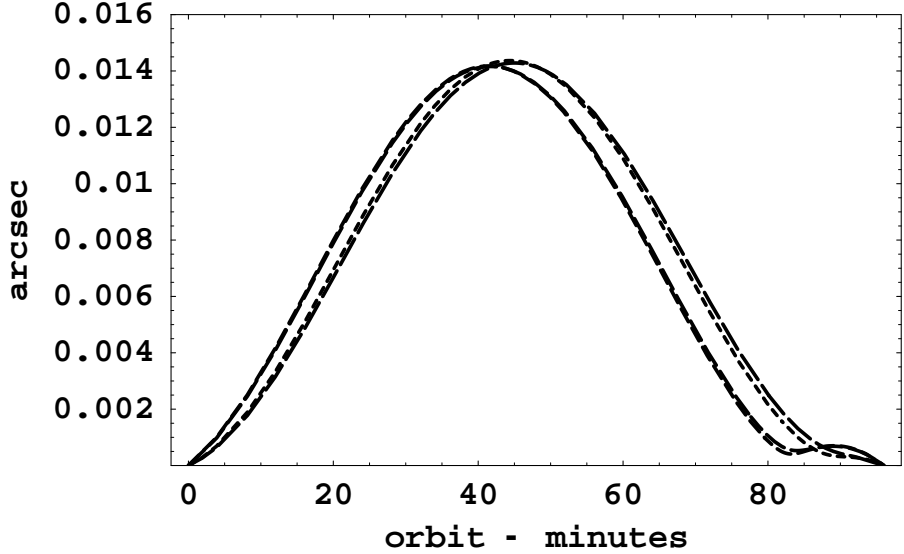}
\caption{\small{ Predicted  variation of the precession angle  $\Delta \Theta=|\Delta {{\bf S}}(t)|/|{\bf S}(0)|$, in
arcsec, over one 97 minute GP-B orbit, from the vorticity induced by the translation of the earth, as given by
(\ref{eqn:precession}).  Predictions are for the  months of April, August, September and February, labeled by 
increasing dash length.  The GP-B expected angle measurement accuracy is 0.0005 arcsec. }
\label{fig:GPB}}\end{figure}

\section{Observing 3-Space Vorticity \label{section:GPB}}
The vorticity effect in (\ref{eqn:equiv10}) can be studied experimentally in the Gravity Probe B (GP-B)
gyroscope satellite experiment in which  
  the precession of four on-board gyroscopes has been measured to unprecedented
accuracy \cite{GPB,Schiff}. In a generalisation of (\ref{eqn:E1})  \cite{Book} the vorticity
$\nabla\times{\bf v}$ is generated by matter in motion through the 3-space, where  here ${\bf v}_R$ is the
absolute velocity of the matter relative to the local 3-space.
\begin{equation}\nabla \times(\nabla\times {\bf v}) =\frac{8\pi G\rho}{c^2}{\bf v}_R,
\label{eqn:CG4b}\end{equation} 
 We then obtain from
(\ref{eqn:CG4b}) the vorticity (ignoring homogeneous vortex solutions) 
\begin{equation}
\vec{\omega}({\bf r},t)
=\frac{2G}{c^2}\int d^3 r^\prime \frac{\rho({\bf r}^\prime,t)}
{|{\bf r}-{\bf r}^\prime|^3}{\bf v}_R({\bf r}^\prime,t)\times({\bf r}-{\bf r}^\prime).
\label{eqn:omega}\end{equation} 

For the smaller earth-rotation induced  vorticity effect
${\bf v}_R({\bf r})={\bf w}\times{\bf r}$ in (\ref{eqn:omega}), where ${\bf w}$ is the angular
velocity of the earth, giving
\begin{equation}
\vec{\omega}({\bf r})_{rot}=4\frac{G}{c^2}\frac{3({\bf r}.{\bf L}){\bf r}-r^2{\bf L}}{2 r^5},
\label{eqn:rotation}\end{equation}
where ${\bf L}$ is the \index{angular momentum - earth} angular momentum of the earth, and ${\bf
r}$ is the distance from the centre.  

In general the vorticity term in 
 (\ref{eqn:equiv10}) leads to an apparent `torque', according to a distant observer, acting on the angular
momentum
${\bf S}$ of the gyroscope,
\begin{equation}
\vec{\tau}= \int d^3 r \rho({\bf r})\; {\bf r}\times(\vec{\omega}({\bf r}) \times{\bf v}_R({\bf r})),
\label{eqn:torque1}\end{equation}
where $\rho$ is its  density, and where now
  ${\bf v}_R$ is used here to describe the motion of the matter forming the gyroscope relative to the local
3-space.  Then
$d{\bf S}=\vec{\tau}dt$ is the change in
${\bf S}$ over the time interval $dt$. For a gyroscope 
${\bf v}_R({\bf r})={\bf s}\times{\bf r}$, where ${\bf s}$ is the angular velocity of the gyroscope.  
This gives
\begin{equation}
\vec{\tau}=\frac{1}{2}\vec{\omega}\times{\bf S}
\label{eqn:torque2}\end{equation}
and so $\vec{\omega}/2$ is the instantaneous angular velocity of precession of the gyroscope, which is thus
equal to the instantaneous angular velocity of 3-space, also relative to a distant observer. The component of
the vorticity in (\ref{eqn:rotation}) has been determined  from the laser-ranged satellites LAGEOS(NASA) and
LAGEOS 2(NASA-ASI)
\cite{Ciufolini}, and the data implies the indicated coefficient on the RHS of (\ref{eqn:CG4b}) to $\pm10\%$.  
For GP-B the direction of ${\bf S}$   has been chosen so that this precession is cumulative and, on averaging 
over an orbit, corresponds to some $7.7\times 10^{-6}$ arcsec per orbit, or 0.042 arcsec per year.  GP-B has
been superbly engineered so that measurements to a precision of 0.0005 arcsec are possible. 

However for the  earth-translation induced precession if we  use $v_R  = 430$ km/s (in the
direction  $\mbox{RA} =5.2^{hr}$, $\mbox{Dec} =-67^0$), (\ref{eqn:omega}) gives
\begin{equation}
\vec{\omega}({\bf r})_{trans}=\frac{2GM}{c^2}\frac{{\bf v}_R\times{\bf r}}{r^3},
\label{eqn:AMomega}\end{equation}
and then the total vorticity  is $\vec{\omega}=\vec{\omega}_{rot}+\vec{\omega}_{trans}$.
The maximum magnitude of the speed of this precession  component is $\omega_{trans}/2=gv_C/c^2=8
\times10^{-6}$arcsec/s,
 where here
$g$ is the usual gravitational acceleration at the altitude of the satellite.   This precession has a different
signature: it  is not cumulative, and is detectable by its variation over each single orbit, as its orbital
average is zero, to first approximation.   Fig.\ref{fig:GPB} shows   $\Delta \Theta=|\Delta {{\bf
S}}(t)|/|{\bf S}(0)|$  over 
 one orbit, where,
\begin{equation}\Delta {{\bf S}}(t) =
\int_0^t dt^\prime \frac{1}{2}\vec{\omega}({\bf r}(t'))_{trans} \times {\bf S}(t^\prime)
 \approx \left(\int_0^t dt^\prime \frac{1}{2}\vec{\omega}({\bf r}(t'))_{trans}\right) \times
{\bf S}(0).
\label{eqn:precession}\end{equation}  
Here $\Delta {{\bf S}}(t)$ is the integrated change in spin, and where
the approximation arises  because the change in
${\bf S}(t^\prime)$ on the RHS of (\ref{eqn:precession}) is negligible.   The plot in  Fig.\ref{fig:GPB}  shows
this effect to be some 30$\times$ larger than the expected GP-B errors, and so easily detectable, if it exists as
predicted herein. 

Essentially then these spin precessions are caused by the rotation of the `wavepackets' describing the matter
forming the  gyroscopes, and caused in turn by the vorticity of 3-space. The above analysis shows that
the rotation is exactly the same as the rotation of the 3-space itself, just as the acceleration of
`matter' was exactly the same as the acceleration of the 3-space. We this obtain a much clearer insight
into the nature of motion, and which was not possible in the spacetime formalism.  

\section{Conclusions\label{section:conclusions}}
We have seen herein that the new theory of 3-space has resulted in a number of fundamental developments, namely that a
complex `quantum foam' dynamical 3-space exists and has a fractal `flow' structure, as revealed most clearly by the
extraordinary DeWitte coaxial-cable experiment. This fractal structure requires that the fundamental equations of
physics  be generalised to take account of, for the first time, the physics of this 3-space and, in
particular, here the inclusion of that dynamics within the dynamics of quantum systems.  We saw that the
generalisation of the Schr\"{o}dinger equation is unique, and that from an Ehrenfest wavepacket analysis we
obtained the equivalence principle, with the acceleration of `matter' being shown to be identical to the
acceleration of the 3-space; which while not unexpected, is derived here for the first time. This result shows
that the equivalence principle is really a quantum-theoretic effect.  As well we obtained by that same
analysis that any vorticity in the 3-space velocity field will result in a corresponding rotation of
wavepackets, and just such an effect is being studied in the GP-B gyroscope experiment.   So for the first time
we see that the original Schr\"{o}dinger equation actually lacked a key dynamical ingredient. As well because 
the 3-space is fractal the generalised Schr\"{o}dinger equation now contains a genuine element of stochasticity.

This research is supported by an Australian Research Council Discovery Grant.

\end{document}